\documentclass[twocolumn,groupedaddress,aps,prx,superscriptaddress,amssymb,amsmath,amsfonts,10pt]{revtex4-2}

\usepackage{bm}% bold math
\usepackage{graphicx}
\usepackage{subfigure}
\usepackage{algorithm}
\usepackage{algpseudocode}
\usepackage{multirow}

\usepackage{wasysym} % permil

\begin{document}

\title{Effect of Stacking Order on the Electronic State of 1\emph{T}-TaS$_2$}

\author{Zongxiu Wu}
\affiliation{Zhejiang Province Key Laboratory of Quantum Technology and Device, Department of Physics, Zhejiang University, Hangzhou 310027, China}
\author{Kunliang Bu}
\affiliation{Zhejiang Province Key Laboratory of Quantum Technology and Device, Department of Physics, Zhejiang University, Hangzhou 310027, China}
\author{Wenhao Zhang}
\affiliation{Zhejiang Province Key Laboratory of Quantum Technology and Device, Department of Physics, Zhejiang University, Hangzhou 310027, China}
\author{Ying Fei}
\affiliation{Zhejiang Province Key Laboratory of Quantum Technology and Device, Department of Physics, Zhejiang University, Hangzhou 310027, China}
\author{Yuan Zheng}
\affiliation{Zhejiang Province Key Laboratory of Quantum Technology and Device, Department of Physics, Zhejiang University, Hangzhou 310027, China}
\author{Jingjing Gao}
\affiliation{Key Laboratory of Materials Physics, Institute of Solid State Physics, Chinese Academy of Sciences, Hefei 230031, China}
\affiliation{University of Science and Technology of China, Hefei 230026, China}
\author{Xuan Luo}
\affiliation{Key Laboratory of Materials Physics, Institute of Solid State Physics, Chinese Academy of Sciences, Hefei 230031, China}
\author{Zheng Liu}
\affiliation{Institute for Advanced Study, Tsinghua University, Beijing 100084, China}
\author{Yu-Ping Sun}
\affiliation{Key Laboratory of Materials Physics, Institute of Solid State Physics, Chinese Academy of Sciences, Hefei 230031, China}
\affiliation{High Magnetic Field Laboratory, Chinese Academy of Sciences, Hefei 230031, China}
\affiliation{Collaborative Innovation Center of Advanced Microstructures, Nanjing University, Nanjing 210093, China}
\author{Yi Yin}
\email{yiyin@zju.edu.cn}
\affiliation{Zhejiang Province Key Laboratory of Quantum Technology and Device, Department of Physics, Zhejiang University, Hangzhou 310027, China}
\affiliation{Collaborative Innovation Center of Advanced Microstructures, Nanjing University, Nanjing 210093, China}

\begin{abstract}
New theoretical proposals and experimental findings on transition metal dichalcogenide 1\emph{T}-TaS$_2$
have revived interest in its possible Mott insulating state.
We perform a comprehensive scanning tunneling microscopy and
spectroscopy experiment on different single-step areas in pristine 1\emph{T}-TaS$_2$.
After accurately determining the relative displacement of the Star-of-David super-lattices in two layers,
we find that different stacking orders can correspond to the similar large-gap spectrum on the upper terrace.
When the measurement is performed away from the step edge, the large-gap spectrum can always be maintained.
The stacking order seems rarely disturb the large-gap spectrum in the ideal bulk material.
We conclude that the large insulating gap is from the single-layer property, which is a correlation-induced Mott gap
based on the single-band Hubbard model.
Specific stacking orders can perturb the state and induce a small-gap or metallic
spectrum for a limited area around the step edge, which we attribute to a surface and edge phenomenon.
Our work provides more evidence about the stacking-order effect on the electronic state and deepens
our understanding of the Mott insulating state in 1\emph{T}-TaS$_2$.
\end{abstract}

\maketitle

\section{Introduction}

The correlation of electrons in many cases plays a crucial role in
determining the special physical properties of quantum
materials~\cite{mott1968metal,imda1998metal,lee2006doping,anderson}.
The layered transition metal dichalcogenide (TMD) 1\emph{T}-TaS$_2$
is a long-known charge density wave (CDW) material with a rich phase
diagram~\cite{manzke1989on,fazekas1979electrical,fazekas1980charge}.
In the commensurate CDW (CCDW) phase, every 13 Ta atoms shrink into a
CDW cluster called star of David (SD) and establish a $\sqrt{13}\times\sqrt{13}$
electronic reconstruction.
The multi-folded band of SD super-lattice induces a flat band around the Fermi energy,
whose narrow bandwidth pushes the low-energy physics to a strong correlation limit.
The conventional view is that each SD
contributes one unpaired electron forming a half-filled flat band
near the Fermi level. With the relatively strong correlation of electrons,
the half-filled band splits into lower and upper Hubbard bands (LHB and UHB),
resulting in a Mott insulating gap~\cite{mott1968metal}. The
insulating state has been identified by many experiments like transport,
scanning tunneling microscopy (STM), angle-resolved photoemission
spectroscopy (ARPES), $\it{etc}$~\cite{fazekas1979electrical,TaS_STM_Mott,TaS_ARPES_Mott}.
Based on this single-band Mott insulator scenario, many exotic phases
like superconducting and quantum spin liquid state have been proposed and explored~\cite{sipos2008from,liu2013superconductivity,Sedoped_SC,QSL,qench_PRR,usr_natphy}.

However, there remain some long-standing unsolved issues about the nature
of the insulating state~\cite{darancet2014three,suzuki2004electronic}.
The density functional theory (DFT) calculation always yields a metallic band
along the $c$-axis in bulk 1\emph{T}-TaS$_2$. It is inconsistent with the
experimental insulating state~\cite{c_resist}.
Further DFT calculations reveal that different stacking orders can change
the electronic state on the $ab$-plane, and even open a gap in the whole Brillouin
zone without considering the onsite-U interaction~\cite{ritschel_natphys,ritschel2018stacking,lee2019origin}.
Experimentally, the high-resolution synchrotron x-ray diffraction (XRD) and ARPES measurements
found some ambiguous evidence of dimerization along the $c$-axis, and simplified
the insulating state into a trivial band insulator~\cite{band_ARPES,dimer_optical}.
Recent STM results revealed
two typical terminations showing insulating gaps with different gap
sizes~\cite{butler2020mottness,double_PRX,Yeom_PRL}.
The large-gap spectrum is similar to the regular spectrum measured in previous experiments.
The large-gap spectrum on the upper terrace corresponds to an AA-stacking order,
in which the SD centers in the upper and lower layers are aligned. The small-gap
spectrum has a much smaller gap size. The small-gap spectrum on the upper terrace
corresponds to an AC-stacking order in which the SD centers in the upper and lower
layers are misaligned. With a unit-cell doubling model, the authors propose that
these two spectra should alternate in successive layers, and they suggest that the
large gap is a band gap originating from the AA-stacked two layers, while the small
gap on a misaligned termination is the single-layer Mott insulating gap.
These DFT and experimental results seem to be compatible, while some fundamental
problems still exist. These results are strongly against the conventional single-band
Mott scenario in 1\emph{T}-TaS$_2$.
Previous experiments show that the large gap can be effectively tuned by external
parameters such as strain, doping, and intercalation
~\cite{sipos2008from,liu2013superconductivity,Sedoped_SC,li2012fe,qiao2017mottness,bu2019possible,Cu_intercalation,yu2015gate},
which cannot be simply explained by the band insulator scenario. The molecular beam epitaxy (MBE)-grown
single-layer TMDs 1\emph{T}-TaS$_2$ and 1\emph{T}-TaSe$_2$ exhibit
a large insulating gap~\cite{lin2020scanning,chen2020strong},
consistent with the most frequently observed large-gap spectrum in bulk 1\emph{T}-TaS$_2$.

To solve the above problems, we implement a systematic and comprehensive STM study
on the stacking order and corresponding electronic state of 1\emph{T}-TaS$_2$.
We report accurate determination of various inter-layer alignments of SD super-lattices and
the associated $\it{dI/dV}$ spectra by exploring single-step areas of the cleaved 1\emph{T}-TaS$_2$.
Through multiple measurements, we find that different stacking orders can correspond to the general
large-gap spectrum in 1\emph{T}-TaS$_2$. The layer-by-layer stacking order is likely to be random
in the bulk material, against a simple inter-layer dimerization mechanism~\cite{lee2019origin,band_ARPES,dimer_optical,butler2020mottness,double_PRX,Yeom_PRL}.
The stacking order seems rarely disturb the large-gap spectrum in the ideal bulk material.
The large gap is therefore most possibly a single-layer feature and  a correlation-driven
Mott gap. Within a limited area around the step edges, stacking orders can
purturb the eletronic state. For the AC- and AB-stacking orders, the top $\it{dI/dV}$
spectra exhibit a small-gap spectrum and a metallic V-shaped spectrum, respectively.
This stacking-order-induced modulation is attributed to a surface and edge phenomenon instead of
the bulk property in 1\emph{T}-TaS$_2$. 
We continue this project in a related work~\cite{zhang2022reconciling}.

\begin{figure}[tp]
\centering
\includegraphics[width=.9\columnwidth]{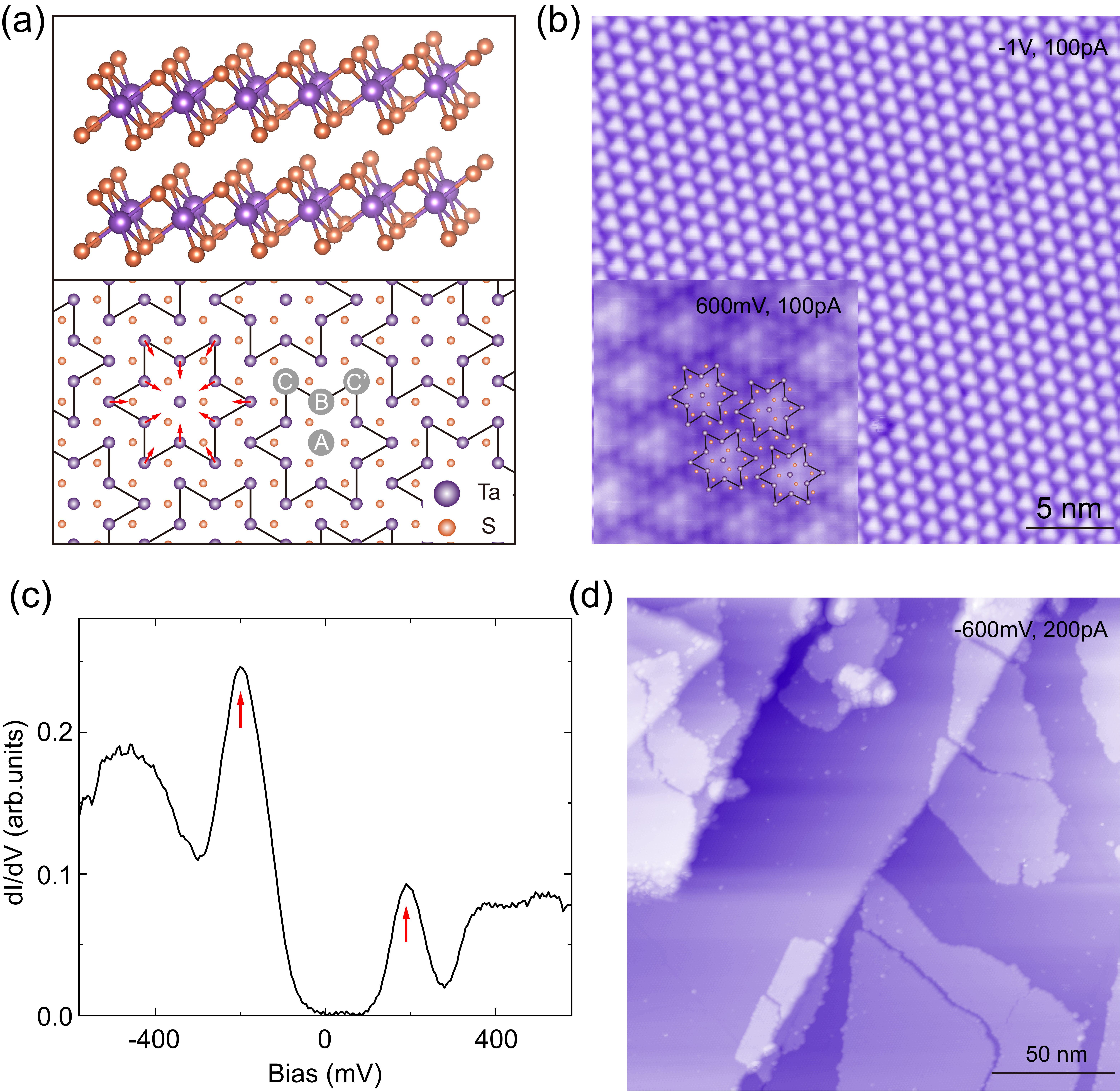}
\caption{Structural and electronic properties of 1\emph{T}-TaS$_2$.
{(a) The upper panel shows the atomic structure of 1\emph{T}-TaS$_2$. The lower panel is a top-view diagram, with the purple and yellow balls representing the Ta atoms and the uppermost S atoms, respectively. Ta atom positions can be divided into central, inner-corner, and outer-corner cases, which are labeled by A, B, and C (C'), respectively. The red arrow indicates the shrinking of the surrounding 12 Ta atoms toward the central Ta. (b) A typical $30\times 30$ nm$^2$ topography in a flat area. The inset displays an atomically resolved topography under the tunneling condition of $V_\mathrm{b}=600$ mV, $I=100$ pA. (c) The typical insulating $\it{dI/dV}$ spectrum of 1\emph{T}-TaS$_2$. The red arrows label the upper and lower Hubbard bands. (d) A complex surface topography.}
}
\label{fig01}
\end{figure}

\section{Experiment and Results}

Each layer of 1\emph{T}-TaS$_2$ is comprised of a triangular lattice of Ta atoms sandwiched
between the top and bottom triangular lattice of S atoms [Fig.~\ref{fig01}(a)]. Each Ta atom is coordinated
octahedrally by S atoms. In the CCDW phase, each SD cluster is formed by 13 Ta atoms, roughly classified
by one central Ta atom (position A), six neighboring Ta atoms at the inner-corner site (position B),
and six next neighboring Ta atoms at the outer-corner site (position C and C').
We glue the sample on a silica substrate or a regular BeCu sheet, cleave the sample at liquid nitrogen (LN$_2$) temperature.
We found little difference between silica and BeCu substrate samples after we performed numerous measurements.
We guess the thermal expansion coefficient of two materials is not large enough to introduce a measurable strain on the surface.
Most cleaved surfaces are very flat as shown in Fig.~\ref{fig01}(b),
while sometimes we can find a surface topography with steps as shown in Fig.~\ref{fig01}(d).
The cleaved sample is thereafter quickly transferred to the STM scan head in the
cryogenic Dewar. We perform the following STM measurement at LN$_2$ temperature.

The SD super-lattice is discerned in the flat surface topography [Fig.~\ref{fig01}(b)],
with each bright spot representing one SD.
The surface exposed S atoms can be
atomically resolved within SDs [inset of Fig.~\ref{fig01}(b)], when the tip is sharp enough.
A typical $\it{dI/dV}$ spectrum
measured at one SD center is shown in Fig.~\ref{fig01}(c).
The two coherence peaks are well established roughly at $\pm 200$ mV.
The resulting insulating gap of 400 meV has been conventionally
considered as a Mott gap between the LHB and UHB.
Peaks outside the Hubbard bands are associated with CDW features.

\begin{figure*}[tp]
\centering
\includegraphics[width=1.7\columnwidth]{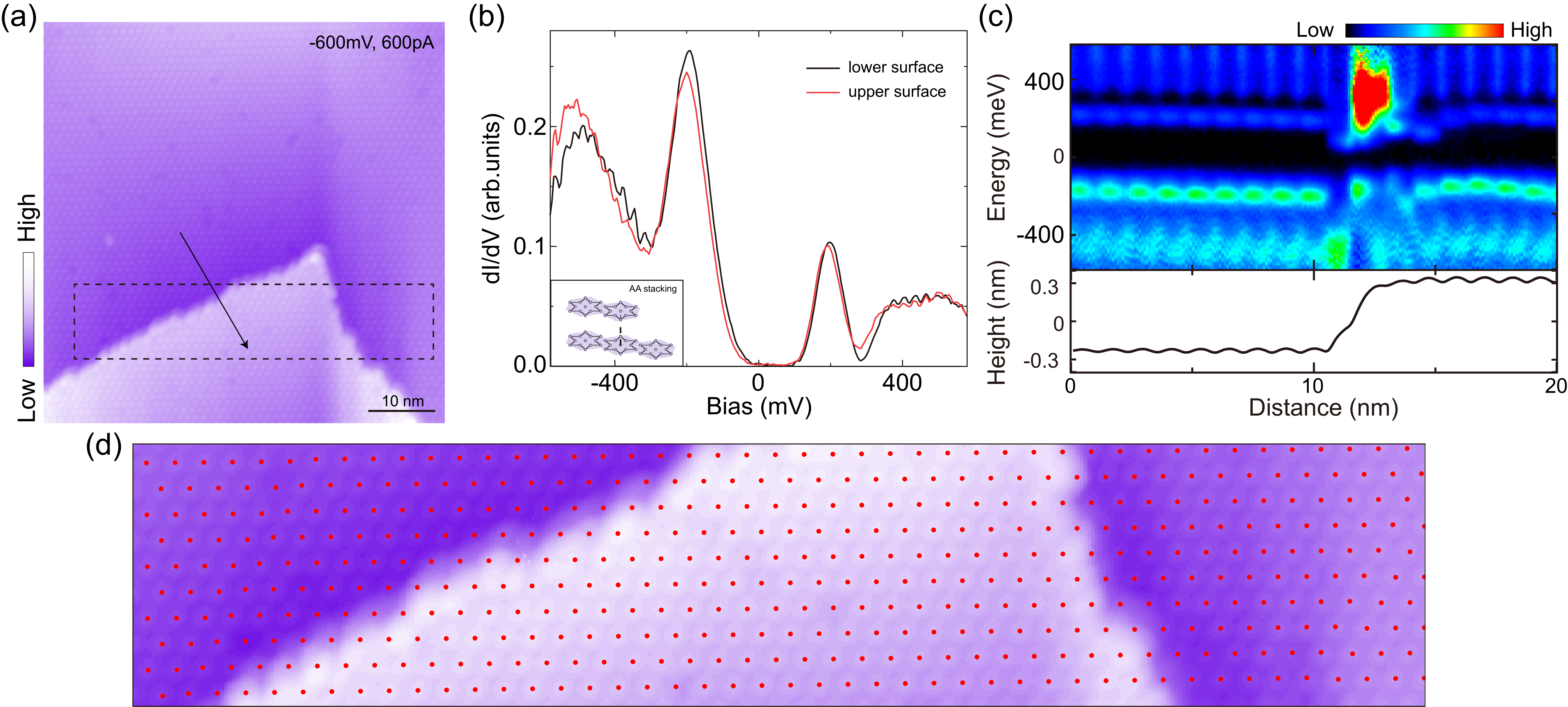}
\caption{AA-stacking single-step area with a top large-gap spectrum.
{(a) STM topography of the single-step area. (b) The red and black curves
show the $\it{dI/dV}$ spectra on the upper and lower terraces, respectively. Both
spectra are similar to the typical spectrum in Fig.~\ref{fig01}(c). The bottom left inset displays a schematic diagram of the AA-stacking order. (c) A series of $\it{dI/dV}$ spectra measured along the black arrowed line in (a), with the height profile displayed in the lower panel. (d) The determination of SD super-lattice for the area within the black dashed frame in (a).}
}
\label{fig02}
\end{figure*}

STM is an ideal tool to determine the relative displacement of SD super-lattices in different layers.
The electronic density of states (DOS) can also be simultaneously
measured to provide correspondence between the stacking order and the electronic state.
Here we specially choose the flat and single-step areas. On the one hand, we measure
a lot of single-step areas and statistically infer the most frequent stacking orders
between neighboring layers in the bulk material. On the other hand, we also
extend the measurement away from the step to check whether the results around the
step edge is compatible with that further away from the step.

Figure~\ref{fig02}(a) shows the topography of a common example of a single-step area.
For both smooth top and underlying layers, the topography includes a triangular
SD super-lattice and each bright spot represents one SD.
Two typical $\it{dI/dV}$ spectra measured on the lower and upper
terraces are displayed in Fig.~\ref{fig02}(b),
both similar to the spectrum in Fig.~\ref{fig01}(c).
To check how the spectrum evolves, a linecut map
is measured along the arrowed line in Fig.~\ref{fig02}(a).
As shown in Fig.~\ref{fig02}(c), the linecut map of $\it{dI/dV}$ spectrum is
vertically displayed in color in the top panel, corresponding to the
height profile in the bottom panel. On both the lower and upper terraces,
the $\it{dI/dV}$ spectrum shows a periodic intensity variation, and shares the similar
large-gap feature except for some modulations at the step edge.
With both the upper and lower terraces exposed, we analyze the
relative displacement of the upper and lower super-lattices.
Within the rectangular area, the red dots in Fig.~\ref{fig02}(d) show the SD array for the lower
layer. We further observe that the red dots at the peninsula coverage area
coincide with the SD centers in the upper layer.
The SD centers in the upper and lower layers are aligned along the $c$-axis.
This stacking order can be defined as the AA-stacking order [inset in Fig.~\ref{fig02}(b)].

\begin{figure}[tp]
    \centering
    \includegraphics[width=0.9\columnwidth]{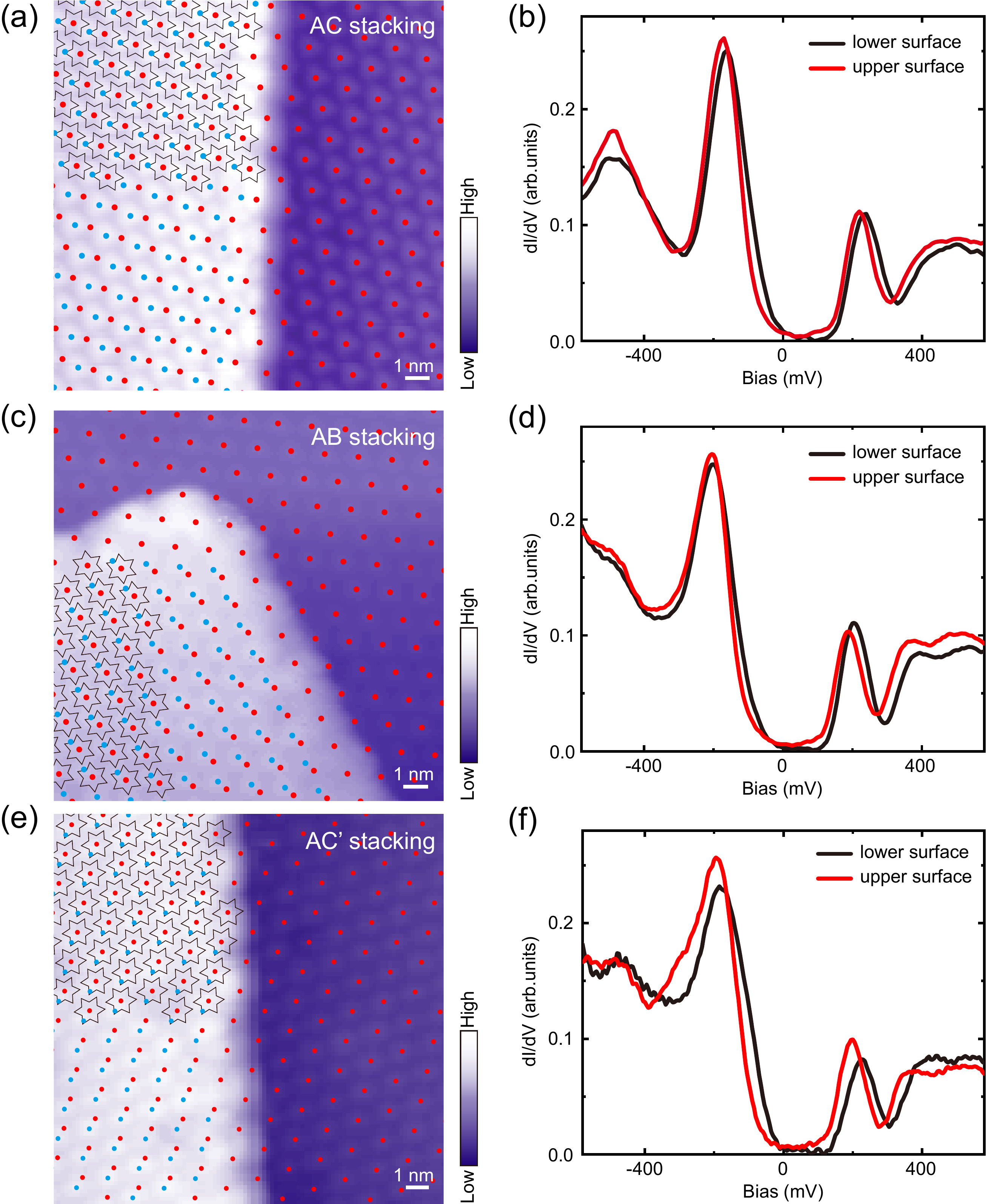}
    \caption{Three misaligned examples on the single-step area.
    {(a), (c), and (e) STM topography of the single-step area. The red (blue) dots label SD centers in the lower (upper) layer,
     showing a misaligned configuration.
    (b), (d), and (f) The red and black curves show the corresponding $\it{dI/dV}$ spectra on the upper and lower terraces in (a), (c), and (e), respectively.
    }
    }
    \label{fig03}
\end{figure}

In the process of collecting and analyzing data, we find examples with misaligned SD super-lattices
in which the large-gap spectrum is obtained on both the upper and lower terraces.
Figure~\ref{fig03}(a), \ref{fig03}(c), and \ref{fig03}(e) display three topographies around different single-step areas.
The SD centers are positioned on both the upper (blue dots)
and lower (red dots) terraces, with red dots also linearly extended to the upper
terrace. The SD centers in the upper
layer are misaligned with those in the lower layer.
Due to the lattice symmetry in 1\emph{T}-TaS$_2$, positioning of SD centers
correspond to two possible atomic arrangements, which differ with a $27.8^{\circ}$ rotation of each SD.
We further apply a specific point defect to obtain the only definite atomic
lattice of the lower layer, with the method shown in Appendix A.
In Fig.~\ref{fig03}(a), the SD centers in the upper layer are aligned with different
but equivalent next nearest Ta atoms in each SD in the lower layer.
This stacking order is defined as the AC-stacking order.
In Fig.~\ref{fig03}(c), the SD centers in the upper layer are aligned with the nearest Ta atoms
in each SD in the lower layer. This stacking order is defined as the
AB-stacking order. The stacking orders in Fig.~\ref{fig03}(e) is determined to be
the AC'-stacking order, with the C' site different from the C site as shown in Fig.~\ref{fig01}(a).
Figure~\ref{fig03}(b), \ref{fig03}(d), and \ref{fig03}(f) display the corresponding $\it{dI/dV}$ spectra measured on
the lower and upper terraces of three topographies, respectively.
They all share the similar large-gap feature in spectrum.
Multiple measurements have been collected in different samples with different tips.
The electronic state and corresponding stacking orders are summarized in Table~\ref{tab01}.
We record 27 times of AA-stacking, 7 times of AC-stacking, 5 times of AB-stacking, and 2 times
of AC'-stacking, all with two large-gap spectra on the lower and upper terraces.
The AA-stacking is relatively dominant, while other stackings also occur a non-negligible
number of times. For all these examples, we also extend the measurement
far away from the step edge, and the large-gap spectrum can
always be maintained (Appendix B). We then presume that the correspondence between the different
stacking orders and the two large-gap spectra is an intrinsic bulk property instead of
just a phenomenon around the step edge.

\begin{table}[]
    \vspace{20pt}
    \centering
    \caption{Counts of stacking orders corresponding to different electronic states.}
    \begin{tabular}{ccc}
        \hline
        Upper surface spectrum & Stacking order & Time \\
       \hline
        Large-gap & AA-stacking & 27\\
         & AC-stacking & 7\\
         & AB-stacking & 5\\
         & AC'-stacking & 2\\
         Small-gap & AC-stacking & 10\\
         Metallic & AB-stacking & 2\\
        \hline
    \end{tabular}
    \label{tab01}
\end{table}

The large insulating gap is the most identified spectrum feature in our experiment.
For all single-step areas, spectrum on the lower terrace exhibits the large-gap spectrum.
For the examples in Fig.~\ref{fig03}, the upper-layer spectrum is not strongly
altered in misaligned stacking conditions.
The consecutive two large-gap spectra on two terraces (Fig.~\ref{fig02}), instead of the alternating large- and small-gap spectra,
rule out the possibility that the large-gap spectrum originates from a unit-cell doubling of
two AA-stacked layers.
When a domain wall appears in the flat area, researchers have always detected two large-gap spectra on both sides of the domain wall~\cite{cho2017correlated}, although different stacking orders are expected to appear on two sides of the domain wall.
Two large-gap spectra corresponding to non-AA stacking orders in Fig.~\ref{fig03} are naturally compatible with the domain wall results.
The non-AA stacking order with consecutive two large-gap spectra further excludes the unit-cell doubling mechanism.
A more reasonable model is that the large insulating gap spectrum reflects the single-layer property,
which is still a Mott insulating gap. Within the ideal bulk material, the stacking order seems to rarely disturb the
large-gap spectrum.

\begin{figure}[tp]
\centering
\includegraphics[width=0.9\columnwidth]{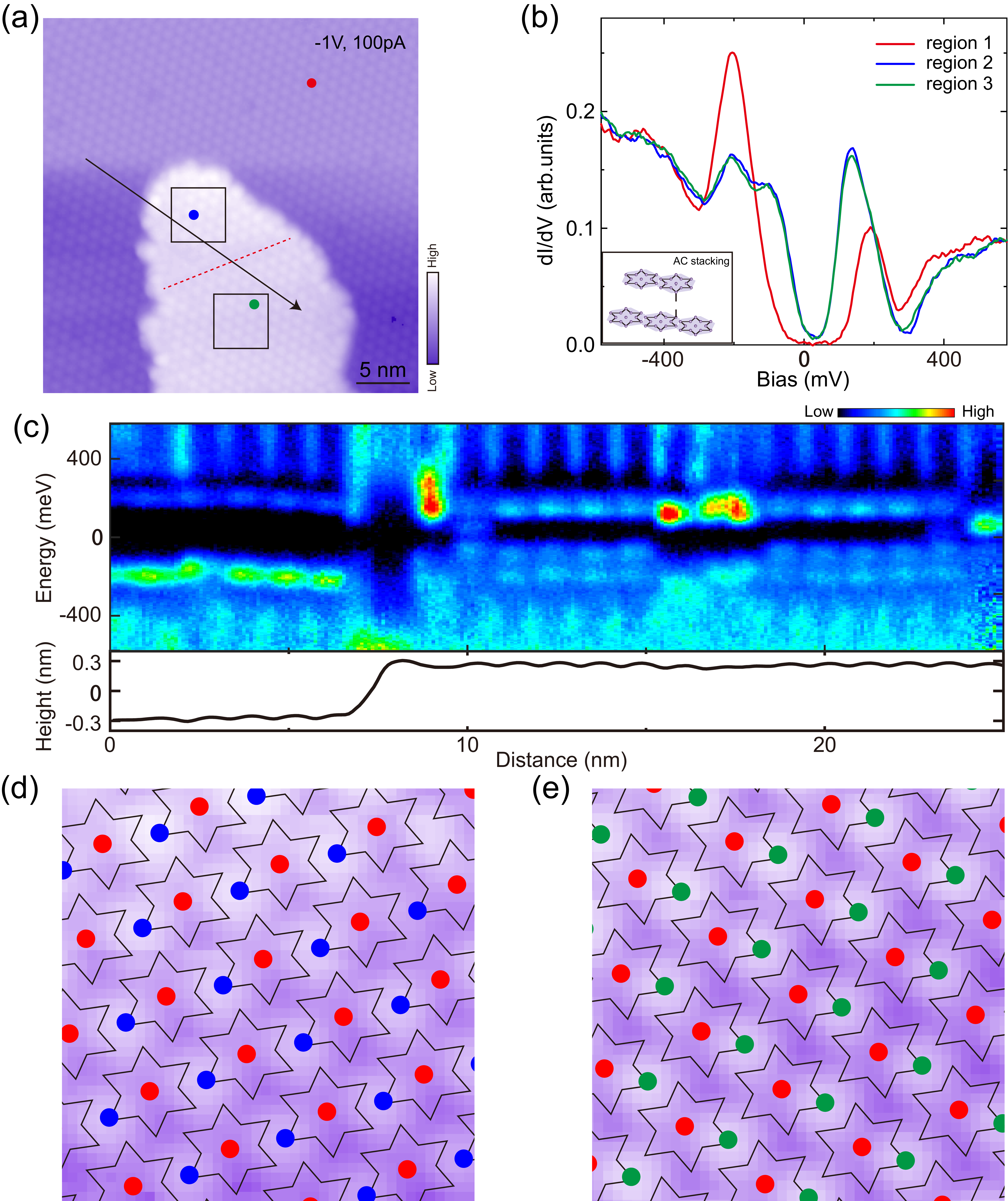}
\caption{AC-stacking single-step area with a top small-gap spectrum.
{(a) STM topography of the single-step area. The red dashed line indicates a domain wall on the top terrace. (b) Three $\it{dI/dV}$ curves correspond to the electronic states measured at the red, blue, and green dot positions in (a), respectively.
(c) A series of $\it{dI/dV}$ spectra measured along the black arrowed line in (a), with the height profile displayed in the lower panel. (d), (e) The enlargement of the two black boxes in (a). The SD centers in the upper layer are all aligned with the SD outer-corner sites in the lower layer.}
}
\label{fig04}
\end{figure}

Besides the large insulating gap, other types of $\it{dI/dV}$ spectra
have been observed on the upper terrace of different single-step areas.
Figure~\ref{fig04}(a) displays a topography with a top peninsula area.
The red dashed line on the upper terrace marks a domain wall.
The SD super-lattices within two domains are shifted relative to each other.
At locations of three points marked in the image, the $\it{dI/dV}$ spectra are measured
on the lower terrace, and two domains on the upper terrace, respectively.
As shown in Fig.~\ref{fig04}(b), the spectrum on the lower terrace is still an
insulating spectrum similar to that in Fig.~\ref{fig01}(c).
In contrast, the spectra measured within both domains on the upper terrace are
different. The UHB peak is shifted to $140$ mV, and the LHB splits into two peaks
at $-100$ and $-205$ mV. Following we call this type of spectrum a small-gap spectrum.
A linecut map is measured along the arrowed line in Fig.~\ref{fig04}(a), with the results
shown in Fig.~\ref{fig04}(c). Together with the periodic intensity variation,
the $\it{dI/dV}$ spectrum shows a large-gap spectrum on the lower terrace and a small-gap spectrum
on the upper terrace, except for some modulations at the step edge and domain wall.

An SD array is extracted for the lower layer, as shown by the red dots in Fig.~\ref{fig04}(d) and \ref{fig04}(e).
The SD centers in the upper layer, labeled by blue and green dots for two domains,
are observed to be not aligned with those in the lower layer.
They are both determined to be the AC-stacking order, as shown in Fig.~\ref{fig04}(d) and ~\ref{fig04}(e).
The similar small insulating gap has been reported with the same stacking
order in recent STM experiments~\cite{butler2020mottness,double_PRX,Yeom_PRL}.
From the periodic intensity variation of $\it{dI/dV}$ spectrum on the upper
terrace [Fig.~\ref{fig04}(c)], we observe that the three shifted peaks around Fermi
energy (zero bias) are still evident at the SD centers, representing that they
are from the Hubbard bands of central Ta orbital~\cite{qiao2017mottness}.
Since we consider the large-gap spectrum a single-layer property, the small-gap spectrum
in our opinion is a perturbed and reduced Mott gap in the AC-stacking
configuration.
During the measurement of single-step areas, we find that the
example in Fig.~\ref{fig04} is also constantly reproducible (10 times in Table~\ref{tab01}).
When we extend the measurements away from the step edge, we find that
the small-gap spectra on the upper terrace often exist within a certain range
of the step edge, e.g. 20 nm (Appendix B). Compared with the AC-stacking case in Fig.~\ref{fig03}(a)-\ref{fig03}(b),
the AC-stacking with a top small-gap spectrum is a more related to a surface and edge
phenomenon.

\begin{figure}[tp]
\centering
\includegraphics[width=0.9\columnwidth]{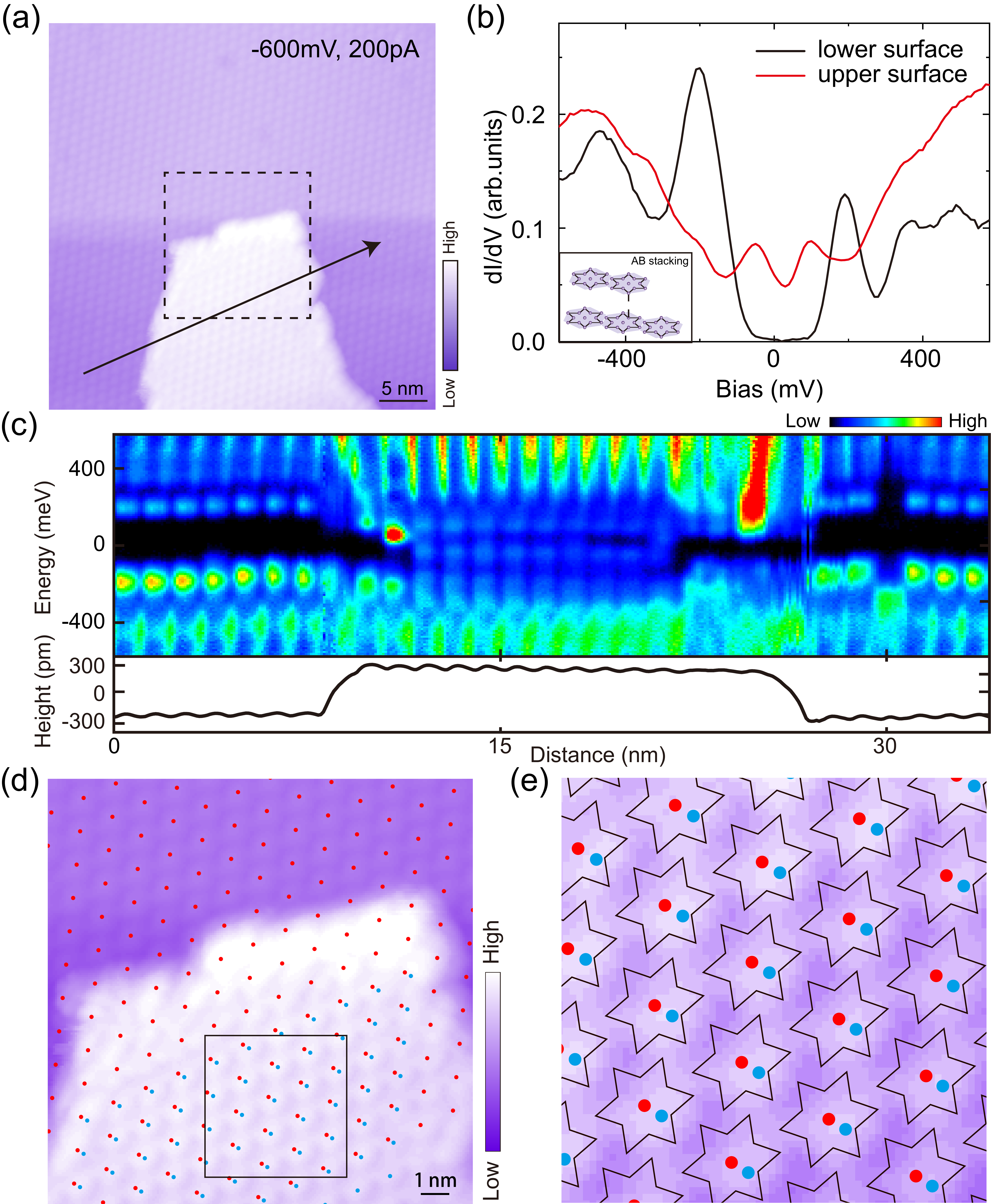}
\caption{AB-type stacking single-step area with a top metallic spectrum.
{(a) The STM topography of a single-step area. (b) The red and black curves show the $\it{dI/dV}$ spectra on the upper and lower terraces, respectively.
(c) A series of $\it{dI/dV}$ spectra measured along the black arrowed line in (a), with the height profile  displayed in the lower panel. (d) The determination of SD super-lattices for the area in the black dashed frame in (a). The red dots label SD centers in the lower layer, and the blue dots label SD centers in the upper layer. (e) The enlargement of the black box in (d). The SD centers in the upper layer are aligned with the SD inner-corner sites in the lower layer. }
}
\label{fig05}
\end{figure}

Another occasional case is presented in Fig.~\ref{fig05} (2 times in Table~\ref{tab01}),
in which the AB-stacking order corresponds to a metallic spectrum on the upper terrace.
From the linecut map measured along the arrowed line [Fig.~\ref{fig05}(c)],
we observe that the coherent peaks of metallic spectrum are evident at the SD centers and originate from
the Hubbard bands of central Ta orbital~\cite{qiao2017mottness}. Although a rare occurrence,
the AB-stacking plays the strongest perturbation to the upper-layer electronic state.
The metallic spectra on the upper terrace exist within a small range of the step edge,
also expected to be a surface or edge phenomenon. The same stacking order can lead to
different electronic states (like Fig.~\ref{fig03}(a) vs. Fig.~\ref{fig04}) means
that some subtle factors other than the stacking order take effect. The stacking-order-induced modulation of
electronic states can also happen under other complex conditions, like the mosaic
state in a pulsed~\cite{ma2016ametallic} or strained~\cite{bu2019possible} sample.
We do not exclude the presence of small-gap or metallic spectra under a
non-ideal local environment within the bulk material.

Taken all the above examples together, we conclude that the examples in Fig.~\ref{fig02} and Fig.~\ref{fig03}
reflect the stacking orders and electronic states in the bulk material. By measuring
a lot of single-step areas, we can infer the most frequent stacking orders
between neighboring layers in the bulk material. The AA-stacking is a prominent
stacking order, while other stacking orders like the AC-stacking also appear from time to
time in our samples. The large-gap spectrum of the single-layer property is, however, not disturbed by
different stacking orders. This picture is compatible with the regular domain wall results in
the bulk material. Around step edges, the AC-stacking and AB-stacking orders can possibly introduce
a modulation of electronic states. Results around the step-edge areas could be surface and edge
phenomena, which should be avoided when we try to infer the intrinsic property in
the bulk material.

We note that all the typical spectra shown in~\cite{butler2020mottness} are also
observed in our experiment, but we provide a different physical picture.
The large insulating gap can
also be well compared with that observed in an MBE-grown single layer
1\emph{T}-TaS$_2$~\cite{lin2020scanning}. Note that although the large-gap
spectrum reflects the single-layer property of 1\emph{T}-TaS$_2$,
the special slope from LHB to zero bias position may be associated with the inter-layer
effect along the $c$-axis. A most recent DFT calculation suggests that the
conventional use of atomic-orbital basis could seriously misevaluate the
electron correlation in the CDW state~\cite{Mott_DFT_PRL}. By adopting a generalized
basis, they successfully reproduce the Mott insulating state under AA-stacking
order condition in 1\emph{T}-TaS$_2$, consistent with our results.

\begin{figure*}[tp]
\centering
\includegraphics[width=1.5\columnwidth]{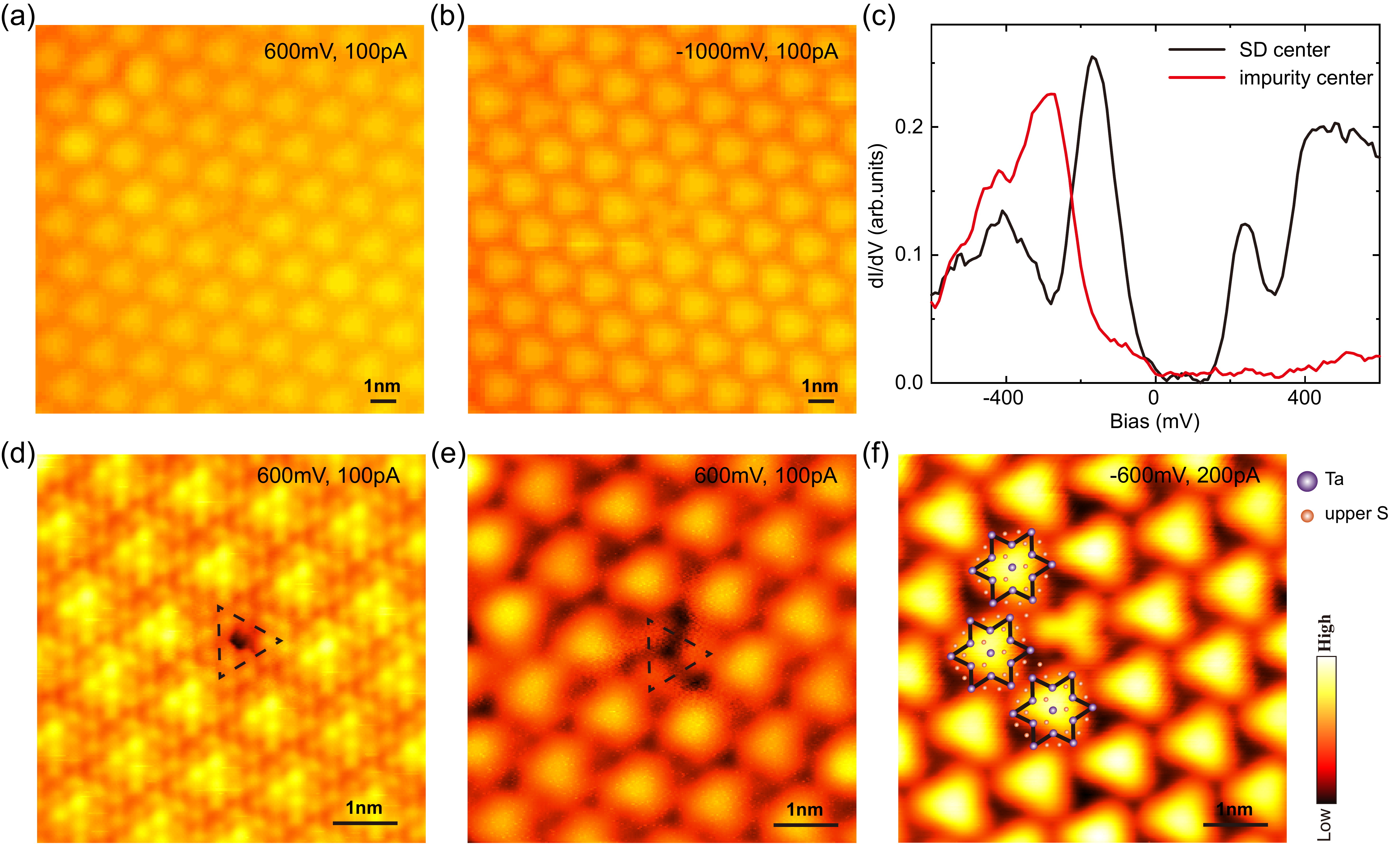}
\caption{Details of the three-petal shaped defect.
{(a), (b) Topographies including the same defect under positive and negative bias voltages. The tunneling condition is $V_\mathrm{b}=-600$ mV and $I=100$ pA in (a), and $V_\mathrm{b}=-1$ V and $I=100$ pA in (b). (c) The red and black curves show the $\it{dI/dV}$ spectra measured at the center of one defect and a normal SD, respectively. (d) An atomically resolved three-petal shaped defect under the tunneling condition of $V_\mathrm{b}=600$ mV and $I=200$ pA. (e) Topography at the same area as in (d). The dashed triangles in (d) and (e) indicate that three petals of the defect are at the sites of three next nearest S atoms. (f) Another topography with a three-petal shaped defect ($V_\mathrm{b}=-600$ mV, $I=200$ pA), with the structure of SD cluster superimposed. }
}
\label{fig06}
\end{figure*}

\section{Conclusion}

In conclusion, our experimental results show that the large insulating gap is a correlation-driven
Mott gap feature of a single-layer 1\emph{T}-TaS$_2$, which is nearly unperturbed in
the bulk material even under different stacking orders. Within a certain range of the step area,
the misaligned stacking orders may disturb the Mott gap to a small-gap or
metallic spectrum, which we attribute to a surface and edge phenomenon.
Further theoretical investigations are required to clarify the subtle interplay
between electron correlation, CDW and stacking order effect.

\begin{acknowledgments}
\noindent{This work was supported by the National Key Research and Development
Program of China (Grants No. 2019YFA0308602),
the Key Research and Development Program of Zhejiang Province, China (2021C01002),
and the Fundamental Research Funds for the Central Universities in China.
Z.L. thanks the National Nature Science Foundation of China (NSFC-11774196) and Tsinghua
University Initiative Scientific Research Program.
J.J.G., X.L., and Y.P.S. thank the support of the National Key Research and Development
Program of China (Grants No. 2016YFA0300404), the National Nature Science Foundation of
China (NSFC-11674326 and NSFC-11874357), the Joint Funds of the National Natural
Science Foundation of China, and the Chinese Academy of Sciences' Large-Scale Scientific
Facility (U1832141, U1932217, U2032215).}
\end{acknowledgments}

\appendix

\section{A Special Type of Point Defect in 1\emph{T}-TaS$_2$}
\label{AppendixA}

\begin{figure}[tp]
\centering
\includegraphics[width=1.0\columnwidth]{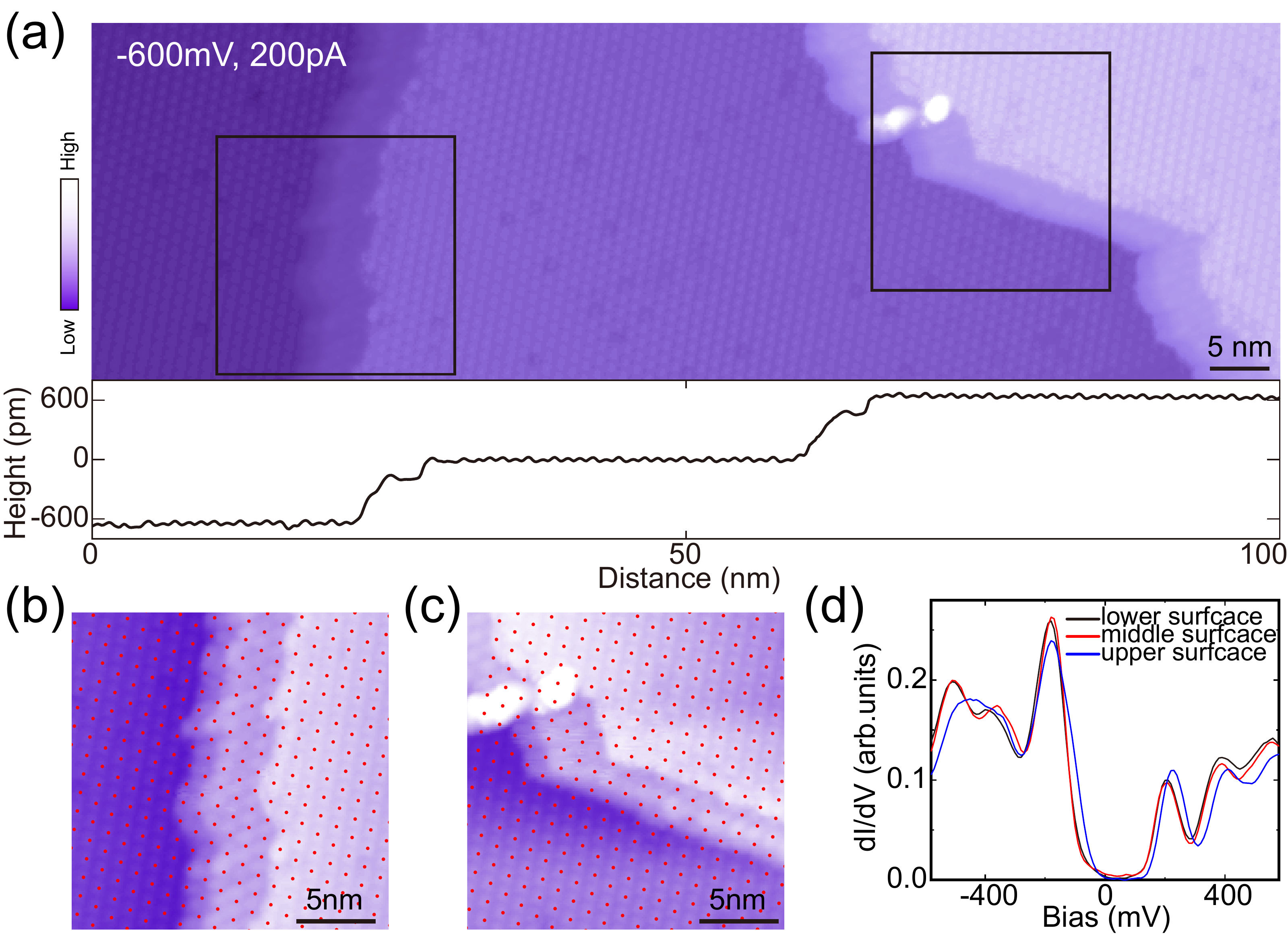}
\caption{Two consecutive AA-stacking steps with a large-gap spectrum.
(a) The STM topography of the two-step area. The single-step edges are blurred as affected by the tip features.
(b) and (c) The determination of SD super-lattice for the left (right) black frame in (a).
(d) The black, red, and blue curves show the dI/dV spectra on the lower, middle, and upper terraces.
}
\label{fig07}
\end{figure}

Figure~\ref{fig06} shows representative data for a particular type of defect we find in 1\emph{T}-TaS$_2$.
Figure~\ref{fig06}(a) and~\ref{fig06}(b) are two topographic images taken at the same area with bias voltage
at 600 mV and -1000 mV, respectively. A three-petal shaped defect can be observed in the middle
of both images, pointing in the same direction but with the three petals in Fig.~\ref{fig06}(b) gathered closer.
We notice that each SD is a triangle in the topography, with the direction
of the triangle in both topographies inverted. Figure~\ref{fig06}(c) shows two $\it{dI/dV}$ spectra measured
at the center of the three-petal defect and a normal SD. Compared with the normal large-gap
spectrum, the defect spectrum shows a totally suppressed UHB while the LHB is left with a
raised peak at -300 mV.

Occasionally, the atomically resolved topography can be obtained when the tip is under a special condition.
Figure~\ref{fig06}(d) and~\ref{fig06}(e) are two topographic images in the same area, with Fig.~\ref{fig06}(d) showing
the atomically resolved surface, on which the top S atoms can be discerned. The brightest atoms
in each SD are the three nearest neighboring S atoms of the central Ta, leading to the
triangle SDs in Fig.~\ref{fig06}(e). The dashed triangles in Fig.~\ref{fig06}(d) and~\ref{fig06}(e) directly illustrate
that the three-petal feature of the defect corresponds to the sites of the next nearest S atoms
of the central Ta atom. In Fig.~\ref{fig06}(d), the black hole within the defect area indicates that
the defect is possibly a missing central Ta, with both UHB and LHB suppressed in the defect spectrum.
The origin of the raised peak below LHB is still an open question at current status.
Then this special type of point defect can help to determine the atomic lattice arrangement as shown in Fig.~\ref{fig06}(f).

\begin{figure}[tp]
\centering
\includegraphics[width=0.9\columnwidth]{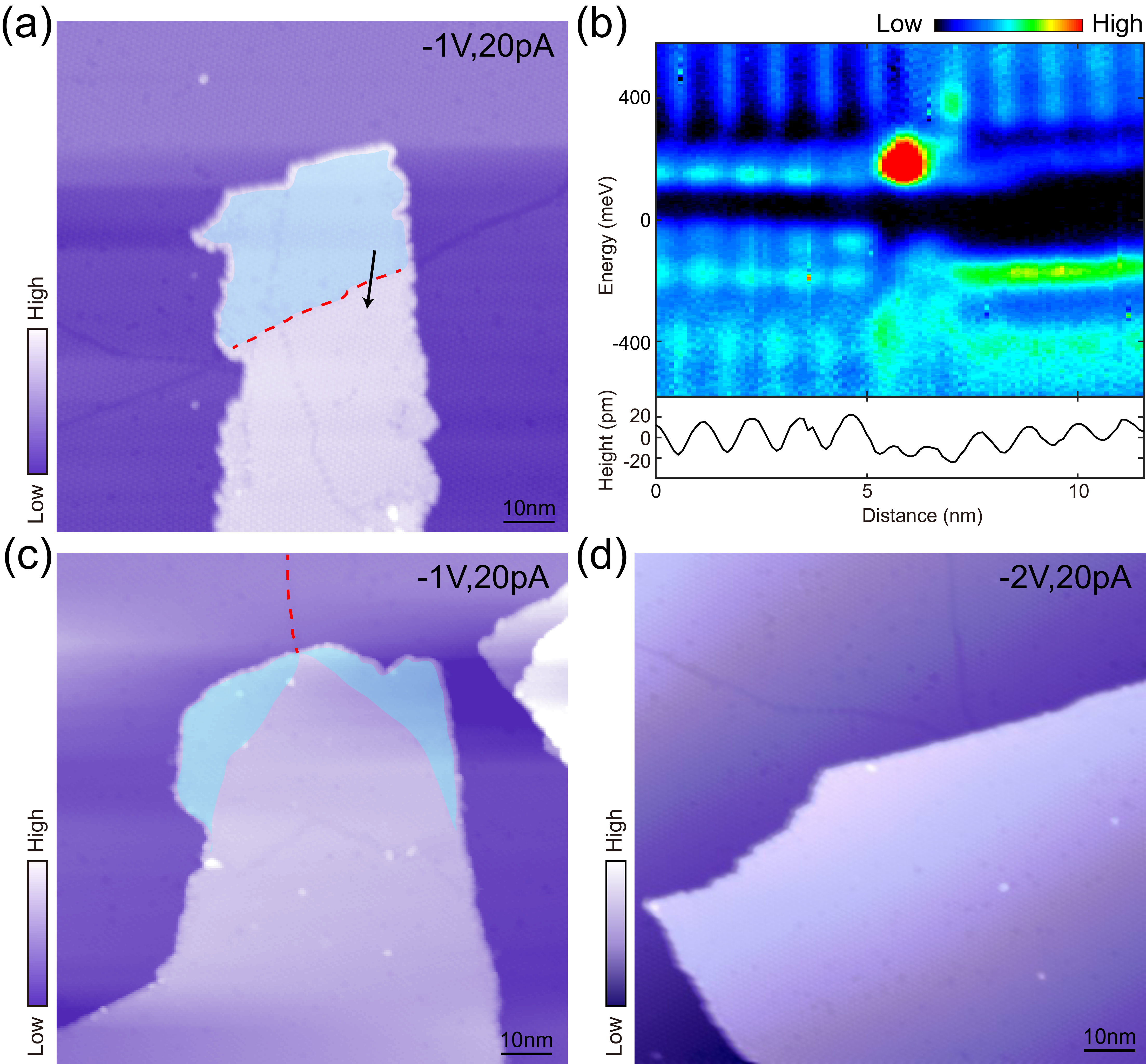}
\caption{Three typical examples of the single-step area results.
(a) and (c) Two single-step area topographies. On the top terrace, small-gap spectra
appear on the blue-shaded areas. Large-gap spectra appear on all the other regions in both topographies.
Red dashed lines indicate domain walls.
(b) A series of dI /dV spectra measured along the black arrowed line in (a), with the height profile
displayed in the lower panel. (d) A single-step area topography. The large-gap spectra appear on the
top terrace and extend the whole area of the top terrace.
}
\label{fig08}
\end{figure}

\section{Different examples of electronic states at step areas}
\label{AppendixB}
In Fig.~\ref{fig07}, we give a two-step area example showing three large-gap spectra compatible with the AA-AA-AA stacking.
Although we mainly apply the single-step area results to infer the stacking order information, the multi-step
result could be compatible with the inferred information. The explicit example of AA-AA-AA stacking can be a
counterexample to the AA-AC-AA-AC stacking in the unit-cell doubling model. Then we further exclude the unit-cell
doubling mechanism.

In Fig.~\ref{fig08}(a) and~\ref{fig08}(c), we show two single-step topographies.
On the top terrace,  the small-gap spectra appear in small areas at the edge of the step (blud shaded areas).
When going away from the step edge, the small-gap spectra recover to the large-gap insulating spectra.
In Fig.~\ref{fig08}(a), this boundary is a domain wall on the top terrace, marked by the red dashed line.
In Fig.~\ref{fig08}(b), we present the $dI/dV$ spectra taken along the black arrowed line in Fig.~\ref{fig08}(a).
On two sides of the domain wall, we could observe an abrupt change from the small-gap spectra to the large-gap spectra.
In Fig.~\ref{fig08}(c), although no domain walls appear on the top terrace, the small-gap spectra also change abruptly
to the large-gap spectra. Guided by the red dashed line of domain wall on the bottom layer, the blue shaded area in
Fig.~\ref{fig08}(c) is possibly surrounded by the bottom-layer domain walls. Figure ~\ref{fig08}(d) shows another
single-step topography. The large-gap spectra appear on the top terrace and extend the whole area of the top terrace.


\begin{thebibliography}{31}
    \bibitem{mott1968metal}
    N. F. Mott,
    \newblock{Metal-Insulator Transition},
    \newblock {Rev. Mod. Phys.}
    \textbf{40}, 677-683 (1968).

    \bibitem{imda1998metal}
    M. Imada, A. Fujimori, and Y. Tokura,
    \newblock{Metal-Insulator Transitions},
    \newblock {Rev. Mod. Phys.}
    \textbf{70}, 1039 (1998).

    \bibitem{lee2006doping}
    P. A. Lee, N. Nagaosa, and X. G. Wen,
    \newblock{Doping a Mott Insulator: Physics of High Temperature Superconductivity},
    \newblock {Rev. Mod. Phys.}
    \textbf{78}, 17-85 (2006).

    \bibitem{anderson}
    D. Belitz and T. R. Kirkpatrick,
    \newblock{The Anderson-Mott Transition},
    \newblock {Rev. Mod. Phys.}
    \textbf{66}, 261-380 (1994).

    \bibitem{manzke1989on}
    R. Manzke, T. Buslaps, B. Pfalzgraf, M. Skibowski, and O. Anderson,
    \newblock{On the Phase Transitions in 1\emph{T}-TaS$_2$},
    \newblock {Europhys. Lett.}
    \textbf{8}, 195-200 (1989).

    \bibitem{fazekas1980charge}
    P. Fazekas and E. Tosatti,
    \newblock{Charge Carrier Localization in Pure and Doped 1\emph{T}-TaS$_2$},
    \newblock {Phys. B + C}
    \textbf{99}, 183-187 (1980).

    \bibitem{fazekas1979electrical}
    P. Fazekas and E. Tosatti,
    \newblock{Electrical, Structural and Magnetic Properties of Pure and Doped 1\emph{T}-TaS$_2$},
    \newblock {Phil. Mag. B}
    \textbf{39}, 229-244 (1979).

    \bibitem{TaS_STM_Mott}
    J. J.Kim, W. Yamaguchi, T. Hasegawa, and K. Kitazawa,
    \newblock{Observation of Mott Localization Gap Using Low Temperature Scanning Tunneling Spectroscopy in Commensurate 1\emph{T}-TaS$_2$}.
    \newblock{Phys. Rev. Lett.}
    \textbf{73}, 2103 (1994).

    \bibitem{TaS_ARPES_Mott}
    F. Zwick, H. Berger, I. Vobornik, G. Margaritondo, L. Forr\'{o}, C. Beeli, M. Onellion, G. Panaccione, A. Taleb-Ibrahimi, and M. Grioni
    \newblock{Spectral Consequences of Broken Phase Coherence in 1\emph{T}-TaS$_2$}
    \newblock{Phys. Rev. Lett.}
    \textbf{81}, 1058 (1998).

    \bibitem{sipos2008from}
    B. Sipos, A. F. Kusmartseva, A. Akrap, H. Berger, L. Forr{\'o}, and E. Tuti{\v{s}},
    \newblock{From Mott State to Superconductivity in 1\emph{T}-TaS$_2$},
    \newblock {Nat. mater.}
    \textbf{7}, 960-965 (2008).

    \bibitem{liu2013superconductivity}
    Y. Liu, R. Ang, W. J. Lu, W. H. Song, L. J. Li, and Y. P. Sun,
    \newblock{Superconductivity Induced by Se-Doping in Layered Charge-Density-Wave System 1\emph{T}-TaS$_{2-x}Se_x$},
    \newblock {Appl. Phys. Lett.}
    \textbf{102}, 192602 (2013).

    \bibitem{Sedoped_SC}
    R. Ang, Y. Miyata, E. Ieki, K. Nakayama, T. Sato, Y. Liu, W. J. Lu, Y. P. Sun, and T. Takahashi,
    \newblock{Superconductivity and Bandwidth-Controlled Mott Metal-Insulator Transition in 1\emph{T}-TaS$_{2-x}Se_x$},
    \newblock {Phys. Rev. B}
    \textbf{88}, 115145 (2013).

    \bibitem{QSL}
    K. T. Law and P. A. Lee,
    \newblock{1\emph{T}-TaS$_2$ as a Quantum Spin Liquid},
    \newblock {Proc. Natl. Acad. Sci. USA}
    \textbf{114}, 6996-7000 (2017).

    \bibitem{qench_PRR}
    H. Murayama, Y. Sato, T. Taniguchi, R. Kurihara, X. Z. Xing, W. Huang, S. Kasahara, Y. Kasahara, I. Kimchi, M. Yoshida, Y. Iwasa, Y. Mizukami, T. Shibauchi, M. Konczykowski, and Y. Matsuda,
    \newblock{Effect of Quenched Disorder on the Quantum Spin Liquid State of the Triangular-Lattice Antiferromagnet 1\emph{T}-TaS$_2$},
    \newblock {Phys. Rev. Research}
    \textbf{2}, 013099 (2020).

    \bibitem{usr_natphy}
    M. Klanj{\v{s}}ek, A. Zorko, R. {\v{Z}}itko, J. Mravlje, Z. Jagli{\v{c}}i{\'c}, P. K. Biswas, P. Prelov{\v{s}}ek, D. Mihailovic, and D. Ar{\v{c}}on,
    \newblock{A High-Temperature Quantum Spin Liquid with Polaron Spins},
    \newblock {Nat. Phys.}
    \textbf{13}, 1130-1134 (2017).

    \bibitem{darancet2014three}
    P. Darancet, A. J. Millis, and C. A. Marianetti,
    \newblock{Three-Dimensional Metallic and Two-Dimensional Insulating Behavior in Octahedral Tantalum Dichalcogenides},
    \newblock {Phys. Rev. B}
    \textbf{90}, 045134 (2014).

    \bibitem{suzuki2004electronic}
    M. T. Suzuki and H. Harima,
    \newblock{Electronic Band Structure of 1\emph{T}-TaS$_2$ in CDW Phase},
    \newblock {J. Magn. Magn. Mater.}
    \textbf{E653}, 272-276 (2004).

    \bibitem{c_resist}
    F. J. Di Salvo and J. E. Graebner,
    \newblock{The Low Temperature Electrical Properties of 1\emph{T}-TaS$_2$},
    \newblock {Solid State Commun.}
    \textbf{23}, 825-828 (1977).

    \bibitem{ritschel_natphys}
    T. Ritschel, J. Trinckauf, K. Koepernik, B. B$\ddot{u}$chner, M. v. Zimmermann, H. Berger, Y. I. Joe, P. Abbamonte and J. Geck,
    \newblock{Orbital Textures and Charge Density Waves in Transition Metal Dichalcogenides},
    \newblock {Nat. Phys.}
    \textbf{11}, 328-331 (2015).

    \bibitem{ritschel2018stacking}
    T. Ritschel, H. Berger, and J. Geck,
    \newblock{Stacking-Driven Gap Formation in Layered 1\emph{T}-TaS$_2$},
    \newblock {Phys. Rev. B}
    \textbf{98}, 195134 (2018).

    \bibitem{lee2019origin}
    S. H. Lee, J. S. Goh, and D. Cho,
    \newblock{Origin of the Insulating Phase and First-Order Metal-Insulator Transition in 1\emph{T}-TaS$_2$},
    \newblock {Phys. Rev. Lett.}
    \textbf{122}, 106404 (2019).

    \bibitem{butler2020mottness}
    C. J. Butler, M. Yoshida, T. Hanaguri, and Y. Iwasa,
    \newblock{Mottness Versus Unit-Cell Doubling as the Driver of the Insulating State in 1\emph{T}-TaS$_2$},
    \newblock {Nat. Commun.}
    \textbf{11}, 2477 (2020).

    \bibitem{double_PRX}
    C. J. Butler, M. Yoshida, T. Hanaguri, and Y. Iwasa,
    \newblock{Doublonlike Excitations and Their Phononic Coupling in a Mott Charge-Density-Wave System},
    \newblock {Phys. Rev. X}
    \textbf{11}, 011059 (2021).

    \bibitem{Yeom_PRL}
    J. Lee, K.-H. Jin, H.W. Yeom,
    \newblock{Distinguishing a Mott Insulator from a Trivial Insulator with Atomic Adsorbates},
    \newblock {Phys. Rev. Lett.}
    \textbf{126}, 196405 (2021).

    \bibitem{band_ARPES}
    Y. D. Wang, W. L. Yao, Z. M. Xin, T. T. Han, Z. G. Wang, L. Chen, C. Cai, Y. Li, and Y. Zhang,
    \newblock{Band Insulator to Mott Insulator Transition in 1\emph{T}-TaS$_2$},
    \newblock {Nat. Commun.}
    \textbf{11}, 4215 (2020).

    \bibitem{dimer_optical}
    Q. Stahl, M. Kusch, F. Heinsch, G. Garbarino, N. Kretzschmar, K. Hanff, K. Rossnagel, J. Geck, and T. Ritschel,
    \newblock{Collapse of Layer Dimerization in the Photo-Induced Hidden State of 1\emph{T}-TaS$_2$},
    \newblock {Nat. Commun.}
    \textbf{11}, 1247 (2020).

    \bibitem{li2012fe}
    L. J. Li, W. J. Lu, X. D. Zhu, L. S. Ling, Z. Qu, and Y. P. Sun,
    \newblock{Fe-Doping-Induced Superconductivity in the Charge-Density-Wave System 1\emph{T}-TaS$_2$},
    \newblock {Europhys. Lett.}
    \textbf{97}, 67005 (2012).

    \bibitem{cho2017correlated}
    D. Cho, G. Gye, J. Lee, S. H. Lee, L. Wang, S. W. Cheong, and H. W. Yeom,
    \newblock{Correlated electronic states at domain walls of a Mott-charge-density-wave insulator 1\emph{T}-TaS$_2$},
    \newblock {Nat. Commun. }
    \textbf{8}, 392 (2017).

    \bibitem{qiao2017mottness}
    S. Qiao, X. T. Li, N. Z. Wang, W. Ruan, C. Ye, P. Cai, Z. Q. Hao, H. Yao, X. H. Chen, J. Wu, Y. Y. Wang, and Z. Liu,
    \newblock{Mottness Collapse in 1\emph{T}-TaS$_{2-x}Se_x$ Transition-Metal Dichalcogenide: an Interplay Between Localized and Itinerant Orbitals},
    \newblock {Phys. Rev. X}
    \textbf{7}, 041054 (2017).

    \bibitem{bu2019possible}
    K. L. Bu, W. H. Zhang, Y. Fei, Z. X. Wu, Y. Zheng, J. J. Gao, X. Luo, Y. P. Sun, and Y. Yin,
    \newblock{Possible Strain Induced Mott Gap Collapse in 1\emph{T}-TaS$_2$},
    \newblock {Commun. Phys.}
    \textbf{2}, 146 (2019).

    \bibitem{Cu_intercalation}
    E. Lahoud, O. N. Meetei, K. B. Chaska, A. Kanigel, and N. Trivedi,
    \newblock{Emergence of a Novel Pseudogap Metallic State in a Disordered 2D Mott Insulator},
    \newblock {Phys. Rev. Lett.}
    \textbf{112}, 206402 (2014).

    \bibitem{yu2015gate}
    Y. J. Yu, F. Y. Yang, X. F. Lu, Y. J. Yan, Y. H. Cho, L. G. Ma, X. H. Niu, S. Kim, Y. W. Son, D. L. Feng, S. Y. Li, S. W. Cheong, X. H. Chen, and Y. B. Zhang,
    \newblock{Gate-Tunable Phase Transitions in Thin Flakes of 1\emph{T}-TaS$_2$},
    \newblock {Nat. Nanotechnol.}
    \textbf{10}, 270-276 (2015).

    \bibitem{lin2020scanning}
    H. C. Lin, W. T. Huang, K. Zhao, S. Qiao, Z. Liu, J. Wu, X. Chen, and S. H. Ji,
    \newblock{Scanning Tunneling Spectroscopic Study of Monolayer 1\emph{T}-TaS$_2$ and 1\emph{T}-TaSe$_2$},
    \newblock {Nano Res.}
    \textbf{13}, 133-137 (2020).

    \bibitem{chen2020strong}
    Y. Chen, W. Ruan, M. Wu, S. J. Tang, H. Ryu, H. Z. Tsai, R. Lee, S. Kahn, F. Liou, C. H. Jia, O. R. Albertini, H. Y. Xiong, T. Jia, Z. Liu, J. A. Sobota, A. Y. Liu, J. E. Moore, Z. X. Shen, S. G. Louie, S. K. Mo, and M. F. Crommie,
    \newblock{Strong Correlations and Orbital Texture in Single-Layer 1\emph{T}-TaSe$_2$},
    \newblock {Nat. Phys.}
    \textbf{16}, 218-224 (2020).

    \bibitem{zhang2022reconciling}
    W. Zhang, Z. Wu, K. Bu, Y. Fei, Y. Zheng, J. Gao, X. Luo, Z. Liu, Y. Sun, and Y. Yin,
    \newblock{Reconciling the bulk metallic and surface insulating state in 1\emph{T}-TaSe$_2$},
    \newblock{Phys. Rev. B}
    \textbf{105}, 035110 (2022).

    \bibitem{ma2016ametallic}
    L. G. Ma, C. Ye, Y. J. Yu, X. F. Lu, X. H. Niu, S. Kim, D. L. Feng, D. Tom\'{a}anek, Y. W. Son, X. H. Chen, and Y. B. Zhang,
    \newblock{A Metallic Mosaic Phase and the Origin of Mott-Insulating State in 1\emph{T}-TaS$_2$},
    \newblock {Nat. Commun.}
    \textbf{7}, 10956 (2016).

    \bibitem{Mott_DFT_PRL}
    D. B. Shin, T. -D. Nicolas, J. Zhang, M. S. Okyay, A. Rubio, and N. Park,
    \newblock{Identification of the Mott Insulating Charge Density Wave State in 1\emph{T}-TaS$_2$},
    \newblock{Phys. Rev. Lett.}
    \textbf{126}, 196406 (2021).




\end{thebibliography}
\end{document}